\documentclass[jgrga]{agu2001} 

\usepackage{graphicx}
\newcommand{\be}{\begin{equation}}
\newcommand{\ee}{\end{equation}}
\newcommand{\ba}{\begin{eqnarray}}
\newcommand{\ea}{\end{eqnarray}}
\authorrunninghead{SORNETTE and WERNER}

\titlerunninghead{APPARENT CLUSTERING AND BACKGROUND EVENTS}

\authoraddr{D. Sornette, Department of Earth and Space Sciences, and
      Institute of Geophysics and Planetary Physics, University of
      California, Los Angeles, California 90095, USA and Laboratoire de
      Physique de la Mati\`ere Condens\'ee, CNRS UMR6622, Universit\'e de
      Nice-Sophia Antipolis, Parc Valrose, 06108 Nice Cedex 2,
      France. (sornette@moho.ess.ucla.edu)}

\authoraddr{M. J. Werner, Department of Earth and Space Sciences, and
      Institute of Geophysics and Planetary Physics, University of
      California, Los Angeles, California 90095, USA. 
(werner@moho.ess.ucla.edu)}

\begin{document}

%
%
%
\setkeys{Gin}{draft=false}
%
%

%
%

\title{Apparent Clustering and Apparent Background Earthquakes Biased
    by Undetected Seismicity}
%

%
%


\author{Didier Sornette}
\affil{Department of Earth and Space Sciences, and Institute of
      Geophysics and Planetary Physics, University of California, Los
      Angeles, USA and Laboratoire de
      Physique de la Mati\`ere Condens\'ee, CNRS UMR6622, Universit\'e de
      Nice-Sophia Antipolis, France}

\author{Maximilian J. Werner}
\affil{Department of Earth and Space Sciences, and Institute of
      Geophysics and Planetary Physics, University of California, Los
      Angeles, USA}

\begin{abstract}

In models of triggered seismicity and in their inversion with empirical data,
the detection threshold $m_d$ is commonly equated to the magnitude $m_0$ of the
smallest triggering earthquake. This unjustified
assumption neglects the possibility of shocks below the detection
threshold triggering observable events. We introduce a formalism
that distinguishes between the detection threshold $m_d$ and the minimum
triggering earthquake $m_0 \leq m_d$. By considering the branching
structure of one complete cascade of triggered events,
we derive the apparent branching ratio $n_a$ (which is
the apparent fraction of aftershocks in a given catalog)
and the apparent background source $S_a$ that are observed when only
the structure above the detection threshold $m_d$ is known
due to the presence of
smaller undetected events that are capable of triggering larger
events. If earthquake triggering
is controlled in large part by the smallest magnitudes as several
recent analyses have shown, this implies
that previous estimates of the clustering parameters
may significantly underestimate the true values: for instance,
an observed fraction of $55\%$ of aftershocks is renormalized
into a true value of $75\%$ of triggered events.

\end{abstract}

%
%

%

\begin{article}

%
%
\section{Introduction}

There are numerous evidences that
a seismic event can have a significant effect
on the pattern of subsequent seismicity,
most obvious in aftershocks of large events.
More recently has emerged an important extension of the concept of
earthquake interactions in the concept of triggered
seismicity, in which
the usual distinction, that foreshocks are precursors of
larger mainshocks which in turn trigger smaller aftershocks, becomes blurred:
a parsimonious and efficient description of seismicity does not
seem to require the division
between foreshocks, mainshocks and aftershocks, which appear indistinguishable
from the point of view of many of their physical and
statistical properties [{\it Helmstetter and Sornette}, 2003a].
An important logical consequence is that cascades
of triggered seismicity (``aftershocks,''  ``aftershocks'' of
``aftershocks,'' ...)
may play an important role in the overall seismicity budget
[{\it Helmstetter and Sornette}, 2003b; {\it Felzer et al.}, 2002].

There is thus a growing interest in phenomenological models of
triggered seismicity,
which use the Omori law as the best coarse-grained proxy for modeling
the complex and multi-faceted interactions between earthquakes,
together with the other most solid stylized facts of seismicity
(clustering in space, the Gutenberg-Richter (GR) earthquake size
distribution and an aftershock productivity law).
This class of ETAS (Epidemic-Type Aftershock Sequences) models
introduced by {\it Ogata} [1988] and {\it Kagan and Knopoff} [1981]
offers a parsimonious approach
replacing the classification of
foreshocks, mainshocks and aftershocks by the concept of earthquake triggering:
earthquakes may trigger other earthquakes through a variety of physical
mechanisms but the effective laws do not allow the identification of a
particular mechanism.

The questions suggested by this approach include:
1) what is the fraction of triggered versus uncorrelated
earthquakes (which is linked to the problem of clustering)? How can one
use this modeling approach to forecast future seismicity? What are the limits
of predictability and how are they sensitive to catalog completeness and
type of tectonic deformation? In general, to attack any such question,
one needs in one way or another to estimate some key parameters
of the models of triggered seismicity.

The royal path is in principle
to use the maximum likelihood method to estimate the
parameters of the considered model from a catalog of
seismicity (time, location and magnitude) (see for instance
{\it Ogata} [1988] and {\it Kagan} [1991]).
The calculation of the likelihood function requires evaluating the theoretical
rate of seismicity at time $t$ induced by all past events at times $t_i<t$.
The maximization of the likelihood with respect to the parameters of the
model, given the data, then provides an estimate of the parameters.
All previous studies have considered that small earthquakes, below the
detection threshold, are negligible. Thus, the rate of
seismicity is calculated as if triggered only by earthquakes
above the detection threshold. However, this method is not correct because
it does not take into account events below the detection threshold,
which may have an important role in the triggering of seismicity.
Indeed, small earthquakes have a
significant contribution in earthquake
triggering because they are much more numerous than larger earthquakes
({\it Helmstetter} [2003]; {\it Felzer et al.} [2002]; {\it
    Helmstetter et al.} [2004]). This can simply be
seen from the competition between
the productivity law $\sim 10^{\alpha M}$ giving the number of events
triggered by a mainshock of magnitude $M$ and the relative abundance
$\sim 10^{-b M}$
of such mainshocks given by the Gutenberg-Richter law: the contribution
of earthquakes of magnitude $M$ to the overall seismic rate is thus
$\sim 10^{-(b-\alpha) M}$, which is dominated by small $M$'s for $\alpha < b$
[{\it Helmstetter}, 2003] or equally contributed by each magnitude
class for $\alpha=b$ ({\it Felzer et al.} [2002]; {\it Helmstetter et
    al.} [2004]). Therefore, one needs to take into account small 
events that are
not observed in order to calibrate correctly models of seismicity and obtain
reliable answers to our questions stated above.
This is an essential bottleneck for the development of
earthquake forecasts based on such models.

The purpose of this note is to present
a general theoretical treatment of the impact of unobserved seismicity
within the framework of models of triggered seismicity. We show by
analyzing the branching structure of a complete cascade (cluster) triggered by
an independent background event
that the unobserved seismicity has the effect of decreasing the
real branching ratio $n$ and of increasing the
number of independent background events $S$ into
apparent quantities $n_a$ and $S_a$. This bias may be very
significant. We therefore claim that previous work should be
reanalyzed from
the new perspective of our approach. This leads also to important consequences
for the methods presently used to forecast future seismicity based only on
incomplete catalogs.

\section{The ETAS model and the smallest triggering earthquake}

\subsection{Definition of the ETAS model}
To make this discussion precise, let us consider the
epidemic-type aftershock sequence (ETAS) model, in which any
earthquake may trigger other earthquakes, which in turn may trigger
more, and so on. Introduced in slightly different forms by {\it Kagan
    and Knopoff} [1981] and {\it Ogata} [1988], the model describes
statistically the spatio-temporal clustering of seismicity.

The triggering process
may be caused by various mechanisms
that either compete or combine, such as pore-pressure changes due
to pore-fluid flows coupled with stress variations, slow redistribution of
stress by aseismic creep, rate-and-state dependent friction within faults,
coupling between the viscoelastic lower crust and the brittle upper crust,
stress-assisted micro-crack corrosion, and more. The ETAS formulation
amounts to a two-scale description: these above physical processes
controlling earthquake interactions
enter in the determination of effective triggering laws in a first step
and the overall seismicity is then seen to result from
the cascade of triggering of events triggering other events
triggering other events
and so on [{\it Helmstetter and Sornette}, 2002].

The ETAS model consists of three assumed laws about the nature of seismicity
viewed as a marked point-process. We restrict this study to the temporal
domain only, summing over the whole spatial domain of interest.
First, the magnitude of any earthquake,
regardless of time, location, or magnitude of the mother shock, is
drawn randomly from the exponential Gutenberg-Richter (GR) law.
Its normalized probability density function (pdf) is expressed as
\be
P(m) = {b \ln(10)10^{-b m} \over 10^{-b m_0}-10^{-b m_{max}}},
~~~~~m_0 \leq m \leq m_{max},
\label{GR}
\ee
where the exponent $b$ is typically close to one, and the cut-offs $m_0$
   and $m_{max}$ serve to normalize the pdf. The upper
cut-off $m_{max}$ is
introduced to avoid unphysical, infinitely large earthquakes. Its
value was estimated to be in the range $8-9.5$ [{\it Kagan}, 1999]. As the
impact of a finite $m_{max}$ is quite weak in the calculations below,
replacing the abrupt cut-off $m_{max}$ by a smooth taper
would introduce negligible corrections to our results.

Second, the model assumes that direct aftershocks are distributed in time
according to the modified ``direct'' Omori law (see {\it Utsu et al.} [1995]
and references therein). Assuming $\theta > 0$, the
normalized pdf of the Omori law can be written as
\be
\Psi(t) = {\theta c^{\theta}\over (t + c)^{1+\theta}} .
\label{pvalue}
\ee

Third, the number of direct aftershocks of an event of magnitude $m$
is assumed to follow the productivity law:
\be
\rho(m) = k ~10^{\alpha (m-m_0)},~~~~~  m_0 \leq m \leq m_{max}.
\label{formrho}
\ee

Note that the productivity law (\ref{formrho}) is zero below
the cut-off $m_0$, i.e. earthquakes smaller than $m_0$ do not trigger
other earthquakes, as is typically assumed in studies using the ETAS
model. The existence of the small magnitude cut-off $m_0$ is
necessary to ensure the
convergence of these types of models of triggered seismicity (in statistical
physics of phase transitions and in particle physics, this is called
an ``ultra-violet'' cut-off which is often necessary to make the
theory convergent). In a closely related paper, {\it Sornette and
    Werner} [2004] showed that the existence of the cut-off $m_0$ has
observable consequences which constrain its physical value.

The key parameter of the ETAS model is defined as the number $n$ of
direct aftershocks per earthquake, averaged over all magnitudes. Here,
we must distinguish between the two cases $\alpha=b$ and $\alpha \neq b$:
\ba
n & & \equiv \int \limits_{m_0}^{m_{max}} P(m) \rho(m) dm
        \nonumber \\
&& = {k b \over b-\alpha} ({1-10^{-(b-\alpha)(m_{max}-m_0)} \over
       1-10^{-b(m_{max}-m_0)}})~~,
\label{nvaqlue}
\ea

for the general case $\alpha \neq b$. The special case $\alpha=b$
gives
\be
n={k b \ln(10) (m_{max}-m_0) \over 1-10^{-b(m_{max}-m_0)}}
\label{n2}
\ee

Three regimes can be distinguished based on the value of $n$. The case
$n<1$ corresponds to the subcritical, stationary regime, where aftershock
sequences die out with probability one. The case $n>1$ describes
unbounded, exponentially growing seismicity [{\it Helmstetter and
Sornette}, 2002].
In addition, the case $b < \alpha$ leads to explosive seismicity with
finite time
singularities [{\it Sornette and Helmstetter}, 2002]. The critical
case $n=1$ separates the two regimes $n<1$ and $n>1$. {\it Helmstetter
   and Sornette} [2003b] showed that the branching ratio $n$ is also
equal to the fraction of triggered events in a seismic catalog.

The fact that we use the same value for the
productivity cut-off and the Gutenberg-Richter (GR) cut-off is not a
restriction as long as the real cut-off for the Gutenberg-Richter law
is smaller than or equal to the cut-off for the productivity law. In
that case, truncating the GR law at the productivity cut-off just
means that all smaller earthquakes, which do not trigger any events,
do not participate in the cascade of triggered events. This should not
be confused with the standard incorrect procedure in many previous
studies of triggered seismicity of simply replacing the GR and
productivity cut-off $m_0$ with the detection threshold $m_d$ in
equations (\ref{GR}) and (\ref{formrho}) (see, for
example, {\it Ogata} [1988]; {\it Kagan} [1991]; {\it Ogata} [1998];
{\it Console et al.} [2003];  {\it Zhuang et al.} [2004]). The
assumption that $m_d=m_0$ may lead to a bias in the estimated parameters.

Without loss of generality, we consider one independent branch
(cluster or cascade of aftershocks set off by a background event) of
the ETAS model. Let thus an independent background event of magnitude $M_1$
occur at some origin of time. The mainshock will trigger direct
aftershocks according to the productivity law (\ref{formrho}). Each of
the direct aftershocks will trigger their own aftershocks, which in
turn produce their own, and so on.
Averaged over all magnitudes, an aftershock produces $n$ direct
offsprings according to (\ref{nvaqlue}). Thus,
in infinite time, we can write the average of the total number
$N_{total}$ of direct and
indirect aftershocks of the initial mainshock as an infinite sum
over terms of (\ref{formrho}) multiplied by $n$ to the power of the
generation [{\it Helmstetter and Sornette}, 2003b], which can be expressed
for $n<1$ as:
\ba
N_{total} & &  = \rho(M_1)+\rho(M_1)~ n+\rho(M_1)~ n^2 + ... \nonumber \\
&& = {k~ ~10^{\alpha (M_1-m_0)} \over 1-n}
\label{Ninf}
\ea
However, since we can only detect events above the detection threshold $m_d$,
the total number of observed aftershocks $N_{obs}$ of the sequence
is simply $N_{total}$ multiplied by the fraction of events above
the detection threshold, given by
\be
f_{obs}={ 10^{b(m_{max}-m_d)}-1 \over 10^{b(m_{max}-m_0)}-1}
\label{Pd}
\ee
according to the GR distribution. The observed number of events in the
sequence is therefore
\ba
N_{obs} & & = N_{total}~ ~f_{obs} \nonumber \\
&& = {k ~~10^{\alpha (M_1-m_0)} \over  1-n}~
\left({10^{b(m_{max}-m_d)}-1 \over 10^{b(m_{max}-m_0)}-1}\right)~.
\label{Nobs}
\ea
Equation (\ref{Nobs}) predicts the average observed number of direct and
indirect aftershocks of a mainshock of magnitude $M_1>m_d$. {\it
    Sornette and Werner} [2004] showed that $m_0$ may be estimated using
fits of $N_{obs}$ given by (\ref{Nobs})
to observed aftershock sequences and B{\aa}th's law. The
essential parameter needed to constrain $m_0$ is the branching ratio
$n$. As we demonstrate below, typical estimates of $n$ in the literature
obtained from a catalog
neglect undetected seismicity and therefore cannot be
used directly to constrain $m_0$.

Naturally, there is no justification for assuming that
$m_d$ should equal $m_0$, as is done routinely in inversions of
catalogs for the parameters of the ETAS model (see, for
example, {\it Ogata} [1988]; {\it Kagan} [1991]; {\it Ogata} [1998];
{\it Console et al.} [2003];  {\it Zhuang et al.} [2004]). First,
detection thresholds change over time as instruments
and network coverage become better, while the physical mechanisms
in the Earth presumably remain the same. No significant deviation from
the Gutenberg-Richter distribution or the productivity law has been
recorded as the detection threshold $m_d$ decreased over time. Second,
studies of earthquake occurrence at small magnitude levels below
the regional network cut-offs show that earthquakes follow the same
Gutenberg-Richter law (for a recent study
of mining-induced seismicity, see, for example, {\it Sellers et al.}
[2003]), while acoustic emission experiments have shown the relevance
of the Omori law at small scales (see, for instance {\it Nechad et
    al.} [2004] and references therein).
Within the assumption of self-similarity, i.e. a continuation of the
GR and productivity laws down to a cut-off, evidence thus points towards
a magnitude of the smallest triggering earthquake and a
Gutenberg-Richter cut-off that lie below the detection threshold and
are thus not directly observable.

\subsection{Two interpretations of the ETAS model}

The ETAS model may be viewed in two mathematically equivalent but
interpretionally different ways. In this section, we develop both
views to underline that our results apply in both cases and to stress
the equivalence of these two views. The first
describes the model as a simple branching model without loops:
The independent background events, due to tectonic loading, may
each independently trigger direct aftershocks, each of which may in
turn trigger secondary shocks, which in turn may trigger more. Because every
triggered event, excluding of course the non-triggered background
events, has exactly one mainshock (mother), but the mother may have
many direct aftershocks (children), the model can
be thought of as a simple branching model without loops. The background
events are assumed to be a stationary Poisson process with a
constant rate. The rate of the aftershocks of a background
event is a non-stationary Poisson process that is updated every
time another aftershock occurs until the cascade dies out. The
intensity is thus conditioned on the specific history of
earthquakes. The expectation of the conditional intensity is an average
over an ensemble of histories. The predicted number of aftershocks of
an independent background event of magnitude $M_1$ as in expression
(\ref{Nobs}) is thus averaged over the ensemble of possible
realizations of the aftershock sequence, and it is also averaged over
all possible magnitudes of the aftershocks. The branching ratio $n$
is therefore an average not only over magnitudes but also over an ensemble
of realizations of the non-stationary Poisson process. The model
thus consists of statistically independent Poisson clusters of events,
which are, however, dependent within one cluster.

The second view of the ETAS model does not allow a unique
identification of the mother or trigger of an earthquake. Rather, each
aftershock was triggered collectively by all previous earthquakes,
each of which contributes a weight determined by the magnitude-dependent
productivity law $\rho(m)$ that decays in time according to the Omori
law $\psi(t)$ and in space according to a spatial function $R(r)$,
often chosen to be an exponential or a power law centered on the
event. The instantaneous conditional intensity rate at some time $t$
at location $r$ is given by
\be
\lambda(t,r)=\mu+\sum_{i|t_i<t}~ \rho(m_i)~\psi(t-t_i)~R(r-r_i)
\label{ETAS}
\ee
where the sum runs over all previous events $i$ with magnitude $m_i$
  at time $t_i$ at location $r_i$.
  Thus the triggering contribution of a
previous event to a later event at time t is given by its own weight
  (its specific entry in the sum)
divided by the total seismicity rate, including the background
rate. A non-zero background rate then contributes evenly to all events
and corresponds to an omnipresent loading contribution. In this way,
earthquakes are seen to be the result of all previous activity
including the background rate. This corresponds to a branching
model in which every earthquake links to all subsequent earthquakes
weighted according to the contribution to triggering. A branching
ratio can then be interpreted as a contribution of a past earthquake
to a future earthquake, averaged over an ensemble of realizations and
all magnitudes. This second view becomes the only possible one for
nonlinear models whose triggering functions depend nonlinearly on
previous events (see e.g. the recently introduced multi-fractal
earthquake triggering model by {\it Ouillon and Sornette} [2004] and
references therein).

These two views are equivalent because the linear formulation
of the seismic rate of the ETAS model together with the exponential
Poisson process ensures that the statistical properties of the
resulting earthquake catalogs are the same. The linear sum over
the individual contributions and the Poisson process formulation are
the key ingredients that allow the model to be viewed as a simple
branching model.

This duality of thinking about the ETAS model
is reflected in the existence of two simulation codes in the
community, each inspired by one of the two views. A program written
by K. Felzer and Y. Gu (personal communication) calculates the
background events as a stationary Poisson process and then simulates
each cascade independently of the other branches as a non-stationary
process. The second code by {\it Ogata} [1998], on the other hand,
calculates the overall seismicity at each point in time by summing
over all previous activity. The latter code is significantly slower
because the independence between cascades is not used, and the entire
catalog is modeled as the sum of a stationary and a non-stationary
process. Despite the different approach, both resulting earthquake
catalogs share the same statistical properties and are thus equally
acceptable. In the interpretation of the model as a simple branching
model, the parameter $n$ defined in (\ref{nvaqlue}) would correspond
to a branching ratio, while the view that aftershocks
are triggered collectively by all previous earthquakes and the
background rate would interpret $n$ as an average contribution of a
single earthquake on future earthquakes. The important point is that
the statistical properties are the same.

While the simulation or forward problem is straight-forward when
adopting the view of the ETAS model as a branching model with one
assigned trigger for any aftershock, the inverse problem of
reconstructing the branching structure from a given catalog can at
best be probabilistic. Because aftershocks of one mother cannot be
distinguished from those of another mother except by spatio-temporal
distance, we have no way of choosing
which previous earthquake triggered a particular event, or whether it
is a background event. Rather, we must resort to calculating the
probability of an event at time t to be triggered by any previous event
according to the contribution that the previous event has
at time t compared to the overall intensity at time t. This
probability is of course equal to the weight or triggering
contribution that a previous event has on a subsequent event when
adopting the collective-triggering view. However, the interpretation
remains different since the probability specifies a unique mother in
a fraction of many realizations.

Having determined from catalogs a branching structure weighted
according to the probability of triggering, one may of course choose to
always pick as source of an
event the most probable contributor, be that a previous event or the
background rate. Another option is to choose randomly according to the
probability distribution and thus reconstruct one possible branching
structure among the ensemble of many other possible ones. The latter
approach has been used by {\it Zhuang et al.} [2004] and
labeled stochastic reconstruction.

The key point is that equating the detection threshold with the smallest
triggering earthquake will most likely lead to a bias in the recovered
parameters of a maximum likelihood analysis as performed by {\it
    Zhuang et al.} [2004] and in many other studies. Therefore, the
weights or probabilities of previous events triggering subsequent
events were calculated from biased parameters.

In the following, we show that the branching ratio and the background
source events are significantly biased when they are estimated from
the apparent branching structure observed above the detection
threshold $m_d$ instead of the complete tree structure down to
$m_0$. We adopt the view of the simple branching model to make the
derivations more illuminating but all results
can be reinterpreted as contributions in the collective-triggering
view.

\section{The Apparent Branching Structure of the ETAS Model}

\subsection{The apparent branching ratio $n_a$}
Seismic catalogs are usually considered complete above
a threshold $m_d$, which varies as a function of technology and
location. For instance, $m_d \approx 2$ for modern Southern California
catalogs (and for earthquakes not too close in time to a large
mainshock [{\it Kagan}, 2003]). The analysis of the statistics of the Omori
and inverse Omori laws for earthquakes of magnitude down to $3$
({\it Helmstetter} [2003]; {\it Helmstetter and Sornette} [2003a])
suggests that $m_0$ is smaller than the completeness magnitude $m_d$
and is thus not directly observable. Thus, $m_0$ is the size of the
smallest triggering earthquake, which most likely differs
significantly in size from the current detection threshold $m_d$. By
considering the branching structure of the model, we derive the
apparent branching ratio and the apparent background source that are
found if only the observed (detected) part of the ETAS model is
analyzed.

As stated above, the ETAS model consists of statistically
independent Poisson-distributed clusters. Within each cluster, each
shock has exactly one trigger, apart from the initial background
event (mainshock) that sets off the cascade. We restrict this study to
the case $n<1$ for mathematical convenience and because this
range of branching ratios gives rise to statistically
stationary seismic sequences.

Since aftershock clusters are independent of each other, averages of
one cluster are equal to ensemble averages, as nothing but the
inherent stochasticity of the model determines the properties of the
clusters. One cluster consists of one independent background event
(source) and its direct and indirect aftershocks. However, if not all
events of the sequence are detected, then there will appear to be less
direct (and indirect) aftershocks, i.e. the branching ratio
will appear different. Furthermore, some observed events will be
triggered by mother-earthquakes below the detection threshold, resulting in
apparently independent background events.

\begin{figure}
       \centering
       \includegraphics[width=7.5cm]{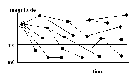}
       \hspace{0.1cm}
       \includegraphics[width=7.5cm]{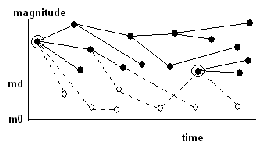}
       \caption{Schematic representations of the branching
structure of the real ETAS model (left) and the apparent
ETAS model (right). The initial mainshock is circled. Only events
above the detection threshold $m_d$ are observed. The apparent
branching ratio does not take into account unobserved triggered
events (dashed lines). An observed event triggered by a mother
below $m_d$ appears as an untriggered background source
event (circled).}
\label{branchETAS}
\end{figure}

We will refer to the observed branching structure above the detection
threshold as the apparent structure.
As a useful visualization of the effect of the detection threshold
on the branching structure, one can think of the branching structure
as a mountain range that is submerged to some fraction
of its height in water (see Figure \ref{branchETAS}).
The height of each mountain corresponds to
the magnitude of an earthquake. To the left (i.e. backward in time) of
each peak is one peak (the trigger) and to the right (i.e. forward in
time) of each peak may be several (the direct aftershocks),
connected from left to right via a ridge linking trigger to offspring.
Since the height of the peaks varies (according to the GR law),
some of the peaks will be below water (below the detection
threshold). A whole part of the branch may be submerged
in water (i.e. unobserved) until eventually a peak rises above the surface and
appears to have no ridge connecting it to
a previous peak, because the ridge of the last observable mountain
simply descends into the water.

This view leads to the conclusion that the average number of
direct aftershocks that are observed will be less than the real
branching ratio,
since some of the triggered events of an observed shock will
fall below $m_d$ and hence not be included in the count. Only the
fraction $f_{obs}$ from equation (\ref{Pd}) above $m_d$ of the total direct
aftershocks $\rho(m)$ will be observed. Moreover, the pdf $P(m|m\geq
m_d)$ of mother
events conditioned on being larger than $m_d$ is zero for $m<m_d$ and
equal to $P(m)/f_{obs}$ for $m_{max}>m\geq m_d$. We can thus define the
apparent branching ratio as
\ba
n_a & & \equiv \int \limits_{m_0}^{m_{max}} P(m|m\geq m_d)~~ \rho(m)~~
f_{obs}~~ dm \nonumber \\
&& = \int \limits_{m_d}^{m_{max}} P(m)~~ \rho(m)~ dm
\\
     && = {k~ b \over b-\alpha}
\left({10^{-(b-\alpha)(m_d-m_0)}-10^{-(b-\alpha)(m_{max}-m_0)} \over
       1-10^{-b(m_{max}-m_0)}}\right) \nonumber
\label{n_a}
\ea

for the case $\alpha \neq b$. The special case $\alpha = b$ gives

\be
n_a = {k b \ln(10) (m_{max}-m_d) \over 1-10^{-b(m_{max}-m_0)}}~~~.
\label{n_a2}
\ee

Using equation (\ref{nvaqlue}) and eliminating $k$, we have $n_a$ in
terms of $n$:
\be
n_a =  n~
\left( {10^{(b-\alpha)(m_{max}-m_d)} -1 \over
      10^{(b-\alpha)(m_{max}-m_0)}-1} \right)~~~~,
\label{n_a_n}
\ee

when $\alpha \neq b$, and

\be
n_a=  n \left( {m_{max}-m_d \over m_{max}-m_0} \right)~~~,
\label{n_a_n2}
\ee

when $\alpha = b$.

\begin{figure}
       \centering
       \includegraphics[width=12cm]{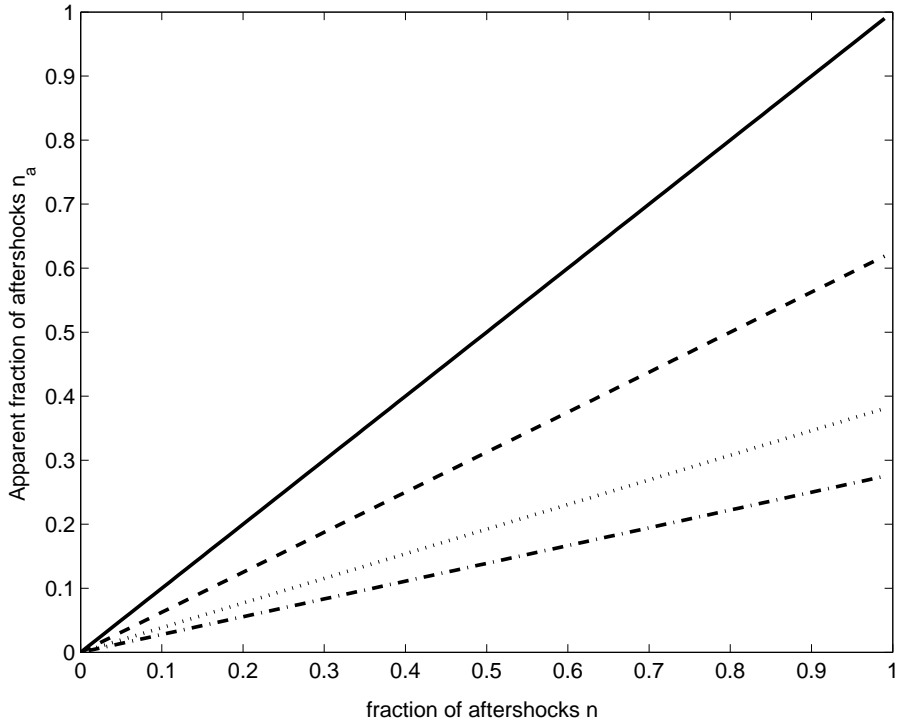}
       \caption{The apparent fraction of aftershocks
       (apparent branching ratio) $n_a$ varies linearly with the real
       fraction of aftershocks (real branching ratio) $n$ with a slope
       fixed by the smallest triggering earthquake $m_0$. As $m_0$
       decreases, the apparent fraction of aftershocks significantly
       underestimates the real fraction. As examples,
       we chose $m_0=m_d=3$ (solid), i.e. $n_a=n$ and no events are
       missed; $m_0=0$ (dashed); $m_0=-5$ (dotted);
       and $m_0=-10$ (dash-dotted). We further assumed
       parameters $m_d=3$, $m_{max}=8$, $b=1$, and $\alpha=1.0$. A small
       value of $\alpha$ amplifies this effect (see Figure \ref{n_a_ratio}).}
\label{fign_a}
\end{figure}

According to expression (\ref{n_a_n}), $n_a \leq n$, where the equality
holds for $m_d$ equal to $m_0$. In principle, equation (\ref{n_a_n})
also holds for $n>1$, but we restrict this study to the stationary
regime $n<1$. Figure \ref{fign_a} shows $n_a$ as a function of $n$
for a range of values of $m_0$ for the case $\alpha=b$. The values of
$m_0$ are $m_0=m_d=3$ (solid), $m_0=0$ (dashed), $m_0=-5$ (dotted) and
$m_0=-10$ (dash-dotted). We assumed $m_d=3$ and $m_{max}=8$. Specifying
$m_0$ then fixes the linear slope of the dependence between $n$ and
$n_a$. Figure \ref{fign_a} demonstrates that the apparent (measurable)
fraction of aftershocks may significantly underestimate the true
fraction of aftershocks even for $m_0$ not very small. For example,
$m_0=-5$ roughly translates a real branching ratio of $n=0.9$ into an
apparent branching ratio $n_a=0.3$. Decreasing $\alpha$ below $b$
places more importance on the triggering from small earthquakes and
therefore strongly amplifies this effect.

\begin{figure}
       \centering
       \includegraphics[width=12cm]{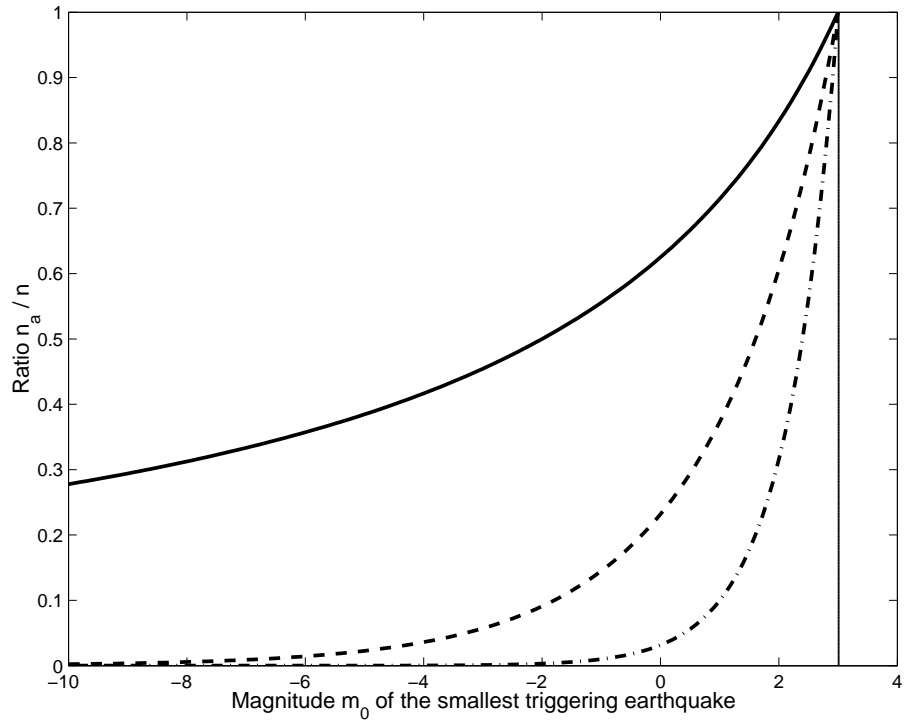}
       \caption{The ratio of the apparent fraction of aftershocks
       (apparent branching ratio) $n_a$ over the real
       fraction of aftershocks (real branching ratio) $n$ varies as a
       function of the smallest triggering earthquake $m_0$. For
       $m_0=m_d$, $n_a=n$ and all events are detected above the
       threshold. For a small value of $m_0$, the ratio becomes small,
       indicating that $n_a$ significantly underestimates
       $n$. Decreasing $\alpha$ amplifies this effect. We
       used parameters $m_d=3$ (vertical reference line), $m_{max}=8$,
       $b=1$. We varied
       $\alpha=0.5$ (dash-dotted), $\alpha=0.8$ (dashed), $\alpha=1.0$
       (solid). }
\label{n_a_ratio}
\end{figure}

In Figure \ref{n_a_ratio}, we plot the ratio $n_a/n$ as a function of
the unknown $m_0$. Again, we assume $m_d=3$, $m_{max}=8$, $b=1$, but
now we let $\alpha=0.5$ (dash-dotted), $\alpha=0.8$ (dashed) and
$\alpha=b=1.0$ (solid). As expected, when $m_0=m_d$, the ratio is one because
there is no unobserved seismicity. As $m_0$ goes to minus infinity,
$n_a$ approaches zero since almost all seismicity occurs below the
threshold. We see clearly that unobserved seismicity
results in a drastic underestimate of the fraction of aftershocks.

Given an estimate of the magnitude of the smallest
triggering earthquake $m_0$ (see {\it Sornette and Werner}
   [2004] and references therein), one can calculate the true branching
ratio from the apparent branching ratio. In fact, {\it Sornette and Werner}
   [2004] obtained four estimates of $m_0$ as a function of $n$ by
comparing the ETAS model prediction of the number of observed
aftershocks (\ref{Nobs}) from fits to observed aftershock sequences
and from the empirical B{\aa}th's law. Their equations (10), (13),
(16) and (18) are the estimates of $m_0$ as a function of $n$ and a
number of known constants specific to the fits to observed aftershock
sequences. We can use these relations of $m_0$ as a function of $n$ to
eliminate $m_0$ from equation (\ref{n_a_n}) to obtain direct estimates
of $n$ as a function of the measurable $n_a$. For simplicity, we
restrict the use of their findings to the case $\alpha=b$. The
estimate resulting from the fits performed by {\it Helmstetter et al.}
[2004] yielded
\be
m_0=m_{max}-({n \over 1-n}){\theta c^{\theta} \over K_{fit}}
{1-10^{-b(m_{max}-m_d)} \over b \ln(10)}
\label{m0_HJK}
\ee
with the values $m_{max}=8.5$, $m_d=3$, $\theta=0.1$, $c=0.001$, $b=1$
and $K_{fit}=0.008$. The study initiated by {\it Felzer et al.} [2002]
provided another estimate
\ba
m_0 & =& m_{max} -{n \over 1-n} {(1-10^{-b(m_{max}-m_d)})\over b
      \ln(10)} \nonumber \\
     && \times {\theta_T c^{\theta_T} \over A_T} 10^{b(M_1-m_d)}~~~~,
\label{m0_F}
\ea
where $m_{max}=8.5$, $m_d=3$, $\theta_T=0.08$, $A_T=0.116$
days$^{-\theta_T}$, $b=\alpha=1$, $c=0.014$ and $M_1=6.04$. Using the
declustering performed by {\it Reasenberg and Jones} [1989],
{\it Sornette and Werner} [2004] obtained
\ba
m_0=m_{max}-{n \over 1-n} {\theta c^{\theta} 10^{-a}\over b
      \ln(10)} (1-10^{-b(m_{max}-m_d)})
\label{m0_RJ}
\ea
where $m_{max}=8.5$, $m_d=3$, $\theta=0.08$, $a=-1.67$, $c=0.05$ and
$b=1$. Finally, using B{\aa}th's law, {\it Sornette and Werner}
[2004] found
\be
m_0= m_{max}-({n \over 1-n}) {(1-10^{-b(m_{max}-m_d)}) \over b
       \ln(10)}~10^{b(M_1-m_a)}
\label{m0_Bath}
\ee
where $M_1-m_a=1.2$ according to the the law, $b=1$, $m_{max}=8.5$,
and $m_d=3$.

\begin{figure}
       \centering
       \includegraphics[width=12cm]{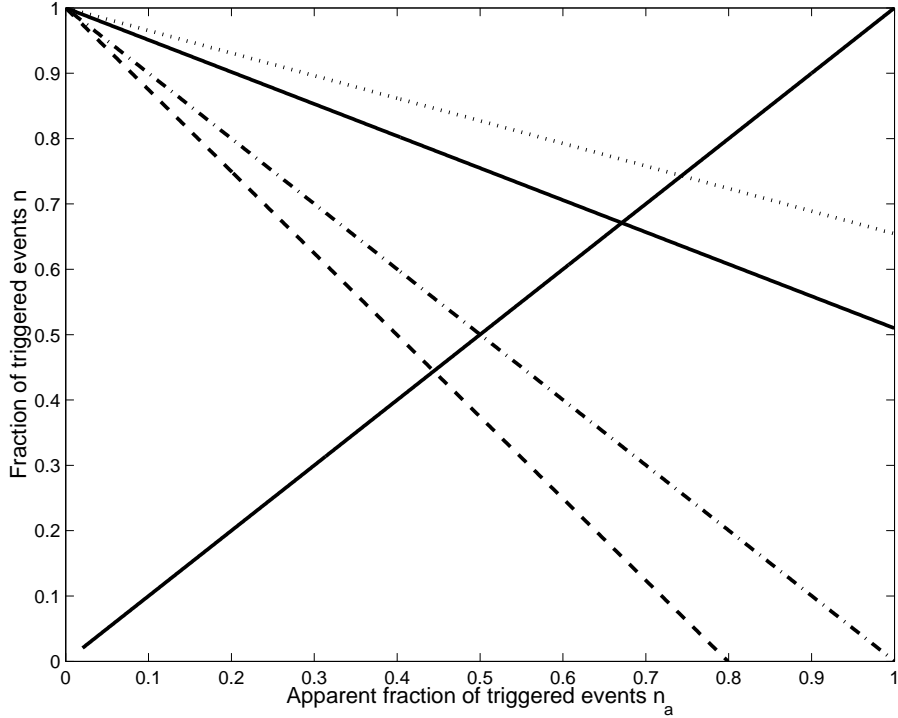}
       \caption{The fraction of aftershocks
       (branching ratio) $n$ can be estimated from the apparent
       fraction of aftershocks (apparent branching ratio) $n_a$ by using
       four estimates of the smallest triggering earthquake $m_0$ as a
       function of $n$ as determined in {\it Sornette and Werner} [2004]
       (see text). The estimates of $m_0$ as a function of $n$ were 
obtained from
       comparisons of the ETAS model
       prediction of the number of observed aftershocks and fits to observed
       aftershock sequences performed by {\it Helmstetter et al.}
       [2004] (solid), {\it Felzer et al.} [2002] (dash-dotted), {\it
       Reasenberg and Jones} [1989] (dotted) and from B{\aa}th's law
       (dashed). The additional diagonal solid line $n_a=n$
       corresponds to $m_0=m_d$ (no undetected events). Along any of the
       four lines, $m_0$ varies from minus infinity to $m_{max}$. Given that we
       can rule out $m_0 \geq m_d$, we can restrict the physical range
       to the left side of the reference curve $n_a=n$. Given an
       estimate of the apparent (measured) fraction of aftershocks
       $n_a$, we can estimate the real fraction from one of the four
       lines. Because $n$ is linearly proportional to $n_a$, an estimate
       of $n_a$ not only determines $n$ from one of the curves, but also
       estimates $m_0$ from the slope of the line connecting the point
       $(n_a,n)$ to the origin. For example, the estimate of {\it
       Helmstetter et al.} [2004] of 55 percent of observed aftershocks
       approximately gives a real fraction $n$ of 75 percent according
       to their own fits to aftershocks and thereby determines $m_0=1.2$
       for $m_{max}=8$ and $m_d=3$.}

\label{fign_a_m0}
\end{figure}

Substituting these four estimates of $m_0$ from equations
(\ref{m0_HJK}), (\ref{m0_F}), (\ref{m0_RJ}), and (\ref{m0_Bath}) into
equation (\ref{n_a_n}) for $n_a$ provides four estimates of $n_a$
versus $n$ all in terms of known constants. These four estimates of
$n$ as a function of $n_a$ can be used to find the correct fraction of
aftershocks from the measurable apparent fraction of
aftershocks. Figure \ref{fign_a_m0} shows these four estimates with
the above constants. As noted above, $n$ varies linearly with $n_a$
with a slope determined by $m_0$. The four estimates of $n$ as a
function of $n_a$ allow for all possible values of $m_0$. The solid
line $n=n_a$, corresponding to the slope $1$ when $m_0=m_d$, separates
the left side of the graph, where $m_0 < m_d$, from the right side,
where $m_0 \geq m_d$, which can be ruled out based on observed
aftershocks from magnitudes $2$. Thus, only the region to the left of
the diagonal should be considered.

Figure \ref{fign_a_m0} can be used to find the real fraction of
aftershocks from the measured apparent fraction by assuming one of the
four estimates of $m_0$ as a function of $n$. For example, {\it
    Helmstetter et al.} [2004] find that $55$ percent of all earthquakes
are aftershocks above $m_d=3$. Using their values to estimate $m_0$ as
a function of $n$, we can determine that the real fraction of
aftershocks is closer to $75$ percent. Thus the size of this effect is
significant. Furthermore, having determined a point on the line
estimating $n$ from $n_a$ for all values of $m_0$ fixes the slope of
$n(n_a)$ and therefore $m_0$. Using their values, we find
$m_0=1.2$. Similar estimates can be made using the apparent fraction
of aftershock values found by {\it Felzer et al.} [2002] and {\it
    Reasenberg and Jones} [1989].

Assuming that current maximum likelihood estimation
methods of the ETAS model parameters, which assume $m_0=m_d$,
determine a branching ratio that corresponds to the present apparent
branching ratio, we can similarly correct these values to find the
true fraction of aftershocks using Figure \ref{fign_a_m0}. For
example, {\it Zhuang et al.} [2004] find a ``criticality parameter'' of
about 45 percent, which we take as a proxy for $n_a$.
Figure \ref{fign_a_m0} shows that the true branching
ratio then lies between $0.45$ and $0.80$, depending on which estimate
(among the four models (\ref{m0_HJK}), (\ref{m0_F}), (\ref{m0_RJ}),
and (\ref{m0_Bath}))
of $m_0$ as a function of $n$ is chosen. These calculations suggest that
previous estimates of the fraction of aftershocks obtained by various
declustering methods significantly underestimated its value.

\subsection{Determination of apparent background events $S_a$ of
uncorrelated seismicity}

In order to derive the number of shocks within one cascade
that are not triggered by a mother above the threshold and thus
appear as independent background events, we need to
distinguish between the case where the initial (main) shock of
magnitude $M_1$ is observable (i.e. $M_1 \geq m_d$) and the case
where it is undetected (i.e. $M_1 < m_d$).

If $M_1 \geq m_d$, then the initial mainshock produces $\rho(M_1) f_{obs}$
observed direct aftershocks. On average, these will
in turn collectively produce $\rho(M_1)~~ f_{obs}~~ n_a$ observed
second generation aftershocks. We specifically do not consider events
above $m_d$ triggered from below $m_d$, which we deal with below in
the definition of the apparent background sources. By continuing
this ``above-water'' cascade for all generations of aftershocks,
we can calculate the number of triggered events that are in direct
lineage above the threshold back to the mainshock as the infinite
sum of terms of $\rho(M_1)~~ f_{obs}$ multiplied by the apparent branching
ratio $n_a$ to the power of the generation. If, on the other hand,
the initial mainshock is below $m_d$, then
no such direct ``above-water'' cascade will be seen.
Any observed shock will be triggered by an event below the water.
Thus, for the two cases, the ``above-water'' sequence is
expressed as:
\be
N_{above} =
\left\{ \begin{array} {r@{\quad \quad}l}
    {\rho(M_1)~f_{obs} \over 1-n_a} = N_{obs} {1-n \over 1-n_a} &
    ,~~M_1 \geq m_d
\\ 0 & ,~~M_1 < m_d \end{array} \right.
\label{Nabove}
\ee
Furthermore, since in the ETAS model, a small earthquake may trigger large
earthquakes, an event below $m_d$ may produce an observed event
above $m_d$. An inversion method that reconstructs the entire
branching structure of the model from an earthquake
catalog will identify these shocks as background events. But
since in reality these events were triggered by earthquakes below the
detection threshold, we will refer to them as apparent background
events. These events can of course trigger their own cascades.
We thus define apparent background source $S_a$ as the number
of observed events above $m_d$ that are apparently not triggered, i.e. have
``mothers'' below $m_d$. Again, we distinguish between the cases
where the mainshock magnitude is $M_1 \geq m_d$ and $M_1 < m_d$. For
the first, $S_a$ is given by the total number of aftershocks below
the threshold multiplied by the average number $r$ of direct aftershocks
they trigger above the threshold. For the second case, we must also
include the direct aftershocks of the initial mainshock that are
observed:
\be
S_a =
\left\{ \begin{array} {r@{\quad \quad}l}
    {\rho(M_1)~ \over 1-n}~(1-f_{obs})~ r & ,~~M_1 \geq m_d
\\{\rho(M_1)~ \over 1-n}~(1-f_{obs})~ r + \rho(M_1)~f_{obs}
& ,~~M_1 < m_d \end{array} \right.
\label{defSa}
\ee
Now, the number $r$ of observable direct aftershocks above $m_d$
averaged over unobserved mothers between $m_0$ and $m_d$ is given
by the following conditional branching ratio:
\ba
r & & \equiv \int \limits_{m_0}^{m_d}~~P(m | m <
m_d)~~\rho(m)~~f_{obs}~~dm
\\ && = (n-n_a)\left({f_{obs} \over 1-f_{obs}}\right)~~~~~,
\label{r}
\ea
where we have used $P(m | m < m_d)= P(m)/(1-f_{obs})$ for $m < m_d$ and
zero otherwise.
Substituting (\ref{r}) into the expression for the apparent source
(\ref{defSa}) and re-arranging using (\ref{Nobs}), we obtain
\ba
S_a & = &
\left\{ \begin{array} {r@{\quad \quad}l}
    {\rho(M_1)~ \over 1-n}~~f_{obs}~~(n-n_a) & ,~~M_1 \geq m_d
\\{\rho(M_1)~ \over 1-n}~~f_{obs}~~(n-n_a) + \rho(M_1)~~f_{obs}~~
& ,~~M_1~<~m_d \end{array} \right. \nonumber\\
& = & \left\{ \begin{array} {r@{\quad \quad}l}
N_{obs}(n-n_a) & ,~~M_1 \geq m_d
\\ N_{obs}(n-n_a) +\rho(M_1)~~f_{obs} & ,~~M_1 < m_d \end{array} \right.
\label{SaN}
\ea

\begin{figure}
       \centering
       \includegraphics[width=12cm]{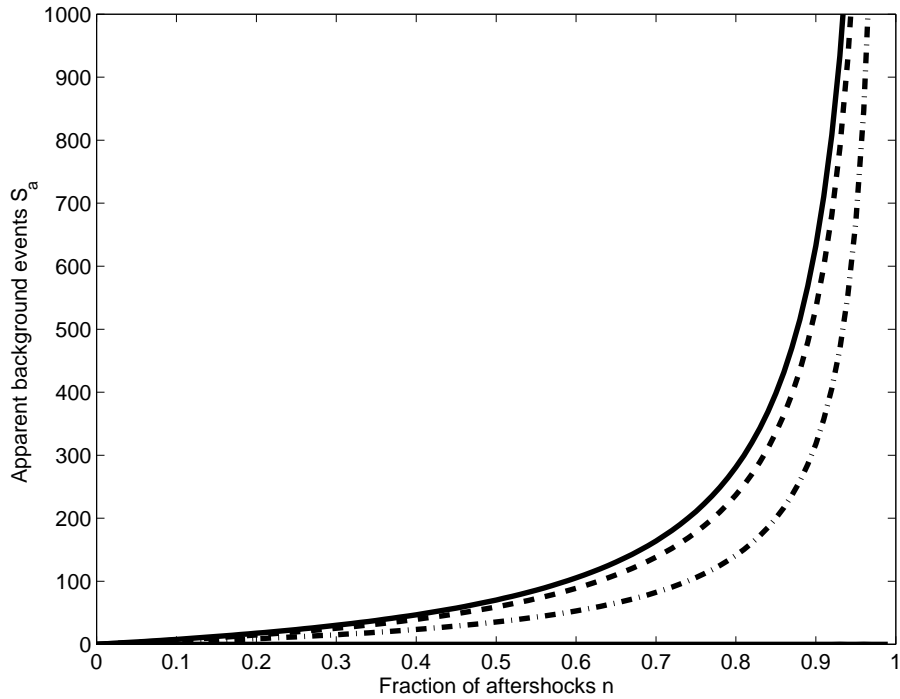}
       \caption{The number of apparent background events $S_a$ in an
       aftershock cascade due to a single background event of magnitude
$M_1=5$ as a
       function of the fraction of aftershocks (branching ratio) $n$ for
       several values of the smallest triggering earthquake. For
       $m_0=m_d$, no events are missed. Therefore the number of apparent
       background events is zero. As $m_0$ decreases, events below the
       detection threshold trigger events above the threshold and
       hence the number of apparent background events increases. We vary
       $m_0=m_d=3$ (solid, coinciding with x-axis), $m_0=0$
       (dash-dotted), $m_0=-5$ (dashed), and $m_0=-10$ (upper solid
       curve). We used parameters $m_d=3$, $m_{max}=8$, $b=1$, and
       $\alpha=1.0$. For very small $m_0$ and $n$ close to $1$, almost all
       events above the detection threshold are triggered from below and
       thus $S_a$ becomes very large (see Figure \ref{sa_ratio}). This
       effect is amplified for decreasing $\alpha$ (not shown).}

\label{sa}
\end{figure}

Equation (\ref{SaN}) shows that, for each genuine background event,
a perfect inversion method would count $S_a$ apparent background events.
Figure \ref{sa} plots the number of apparent background events $S_a$
as a function of the branching ratio $n$ for an example aftershock
cascade set off by a magnitude $m=5$ initial shock. We assumed the
values $m_{max}=8.5$, $m_d=3$ and $\alpha=b=1.0$. The figure shows
that for one cascade, i.e. one independent background event, hundreds of
earthquakes appear as apparent background events when $m_0 < m_d$.

\begin{figure}
       \centering
       \includegraphics[width=12cm]{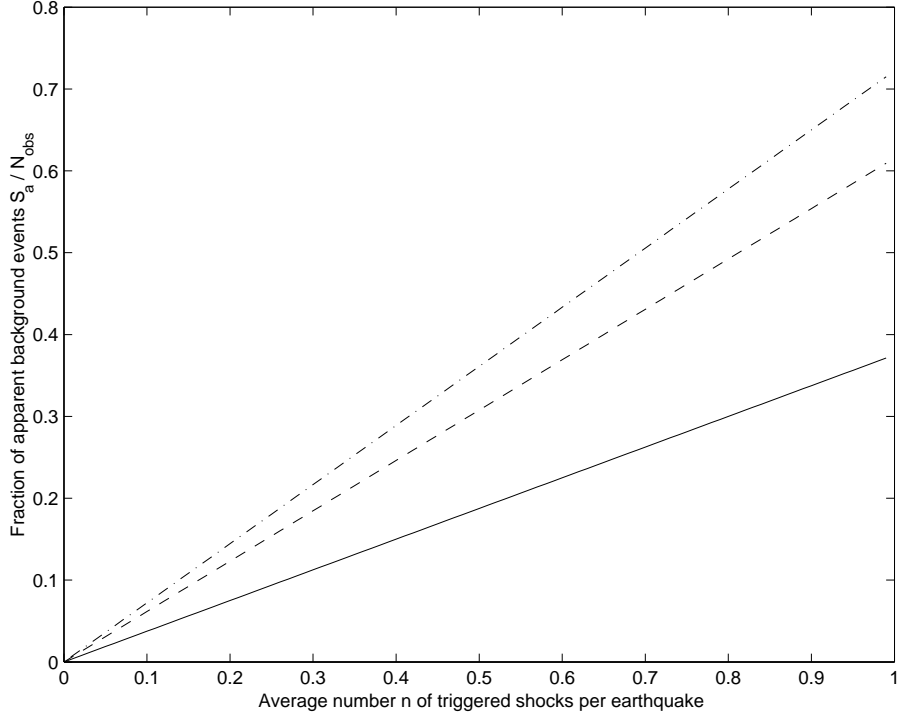}
       \caption{The ratio of the number of apparent background events
       $S_a$ over the total observed number $N_{obs}$ of aftershocks of
       one cascade varies as $n-n_a$. Here, we show the ratio as a
       function of the branching ratio $n$ by assuming a particular
       value of $m_0$. For $m_0=m_d$ (solid, coinciding with x-axis),
       there are no apparent background sources. For $m_0$ less than
       $m_d$, the ratio increases as more and more of the observed
       events are triggered by unobserved events. As examples, we show the
       ratio $S_a/N_{obs}$ for $m_0=m_d=3$ (solid, coinciding with
       x-axis), $m_0=0$ (upper
       solid line), $m_0=-5$ (dashed) and $m_0=-10$ (dash-dotted) as a
       function of the branching ratio $n$ (average number of
       aftershocks per earthquake also equal to the fraction of
       aftershocks in a catalog) for parameters $m_d=3$, $m_{max}=8$,
       $b=1$, and $\alpha=1.0$. For very small $m_0$, $n_a$ approaches
       zero and the ratio $S_a/N_{obs}$ approaches its limiting value
       $n$, meaning that almost all observed earthquakes were triggered
       by events below the detection threshold $m_d$. The effect of
       unobserved events triggering observed quakes resulting in an
       apparent background source rate is further amplified by smaller
       values of $\alpha$ (not shown).}

\label{sa_ratio}
\end{figure}

In Figure \ref{sa_ratio}, we investigate the relative importance of
the apparent background events with respect to the observed number of
aftershocks of one cascade. According to equation (\ref{SaN})
\be
S_a/N_{obs}=n-n_a~,
\ee
i.e. a significant fraction $n-n_a$ of events of the
actually observed aftershocks are falsely identified as background
events (since all events are really triggered from a single
mainshock in our example). For $m_d=m_0$, the ratio is zero, since no
events trigger
below the detection threshold. However, as $m_0$ decreases and more
and more events fall below $m_d$, the fraction increases until $n_a$
goes to zero and the ratio approaches $n$. This effect increases with
decreasing $\alpha$. Small values of $\alpha$ generally place more
importance on the cumulative triggering of small earthquakes.

\subsection{Consistency check: $N_{obs}$ as the sum of
``above-water'' cascades triggered by the mainshock
and by the apparent background events}

To complete the calculations and show consistency of the results, we
now demonstrate that the observed cascades set off by the apparent background
events, when added to the original ``above-water'' cascade, add up to
the total observed number of aftershocks of the whole sequence. Each
apparent source event will trigger its own cascade above the threshold
$m_d$ with branching ratio $n_a$. The total number of events due to
the apparent background events and their cascades above the threshold
is
\be
N_{source}= S_a + S_a ~ n_a + S_a ~ n_a^2 + ...
= {S_a \over 1-n_a}~.
\label{Nsourcepre}
\ee
Substituting expression (\ref{SaN}) and using (\ref{Nobs}) gives
\be
N_{source} =
\left\{ \begin{array} {r@{\quad \quad}l}
      N_{obs}~{(n-n_a) \over 1-n_a} & ,~~M_1 \geq m_d
\\ N_{obs}~{(n-n_a) \over 1-n_a} + {\rho(M_1-m_0)~f_{obs} \over 1-n_a}
& ,~~M_1<m_d \end{array} \right.
\label{Nsource}
\ee
Combining the direct ``above-water'' cascade (\ref{Nabove}) with
the apparent source cascades (\ref{Nsource}) gives the total amount
of apparent events observed after the initial event
\ba
N_a & = & N_{source} + N_{above} \nonumber\\
& = & \left\{ \begin{array} {r@{\quad \quad}l}
      N_{obs}~{(n-n_a) \over 1-n_a} + {\rho(M_1-m_0)~f_{obs} \over 1-n_a} &
      ,~~M_1 \geq m_d \\
N_{obs}~{(n-n_a) \over 1-n_a} + {\rho(M_1-m_0)~f_{obs}
        \over 1-n_a} + 0  & ,~~M_1 < m_d \end{array} \right.~\nonumber \\
&=& N_{obs},
\label{Napp}
\ea
where $N_{obs}$ is given by (\ref{Nobs}).
The last equality confirms the consistency of our decomposition
into apparently-triggered earthquakes and apparent sources.

Expressions (\ref{n_a}) and (\ref{SaN}) show that analyzing the tree structure
of triggered seismicity only above the detection threshold leads to
the introduction of an apparent source $S_a$ and an apparent
branching ratio $n_a$. It is important to realize that both
are renormalized simultaneously by $m_d \neq m_0$.
If the current inversion techniques for the ETAS
parameters were perfect and correctly reconstructed the
tree structure of all sequences, the inverted values would
be equal to our analytical results (\ref{n_a}) and (\ref{SaN}).
Accordingly, the value
of the background source would be overestimated and
the branching ratio underestimated. In fact, one single true
sequence will appear as many different sequences, each apparently set
off by an apparent background event.

\section{Conclusions}

We have shown that unbiased estimates of the fraction of aftershocks and the
number of independent background events are simultaneously renormalized to
apparent values when the smallest triggering earthquake $m_0$ is smaller
than the detection threshold $m_d$. In summary, mainshocks above the
threshold will
appear to have less aftershocks, resulting in a smaller apparent
branching ratio. Meanwhile, unobserved events can trigger events above
the threshold giving rise to apparently independent background events
that seem to increase the constant background rate to an apparent
rate. Assuming that current techniques which are used to
invert for the parameters of the ETAS model (for example, the maximum
likelihood method) under the assumption $m_d=m_0$ are unbiased
estimators of $n_a$ and $S_a$, then the obtained values for the fraction
of aftershocks and the background source rate correspond to renormalized
values because of the assumption that the detection threshold $m_d$
equals the smallest triggering earthquake $m_0$. We predict that $n$ will be
drastically underestimated and $S$ strongly overestimated for $m_0$
much smaller than $m_d$.

\begin{acknowledgments}
We acknowledge useful discussions with A. Helmstetter and J. Zhuang.
This work is partially supported by NSF-EAR02-30429, and by
the Southern California Earthquake Center (SCEC). SCEC is funded by
NSF Cooperative
Agreement EAR-0106924 and USGS Cooperative Agreement 02HQAG0008.  The
SCEC contribution number for this paper is xxx. MJW gratefully
acknowledges financial support from a NASA Earth System Science
Graduate Student Fellowship.
\end{acknowledgments}

%
%
%
%
%
%
%
%


\end{article}

\newpage

%
%
%

\newpage

\newpage

\newpage

\newpage

\newpage

\end{document}